# A tutorial on electrogastrography using low-cost hardware and open-source software

Evgeniya Anisimova* [1], Sameer N.B. Alladin* [1], Styliani Tsamaz [1], & Edwin S. Dalmaijer [1]

** Authors contributed equally.*

1. School of Psychological Science, University of Bristol, United Kingdom

**Corresponding Author**

Dr Edwin S. Dalmaijer, School of Psychological Science, University of Bristol, 12a Priory Road, Bristol, BS8 1TU, United Kingdom. edwin.dalmaijer@bristol.ac.uk

**Author contributions (CReditT)**

Conceptualisation: ESD; data curation: SNBA, ESD; formal analysis: SNBA, EA, ESD; investigation: EA, ST; methodology: EA, SNBA, ESD; software: ESD; supervision: ESD, visualisation: EA, ESD; writing (original draft): EA, SNBA; writing (review and editing): EA, SNBA, ST, ESD.

**Acknowledgements**

During data collection for this project, EA was enrolled in the *MSci in Psychology and Neuroscience* and ST in the *BSc in Psychology* at the University of Bristol.

## Abstract

Electrogastrography is the recording of changes in electric potential caused by the stomach's pacemaker region, typically through several cutaneous sensors placed on the abdomen. It is a worthwhile technique in medical and psychological research, but also relatively niche. Here we present a tutorial on the acquisition and analysis of the human electrogastrogram. Because dedicated equipment and software can be prohibitively expensive, we demonstrate how data can be acquired using a low-cost OpenBCI Ganglion amplifier. We also present a processing pipeline that minimises attrition, which is particularly helpful for low-cost equipment but also applicable to top-of-the-line hardware. Our approach comprises outlier rejection, frequency filtering, movement filtering, and noise reduction using independent component analysis. Where traditional approaches include a subjective step in which only one channel is manually selected for further analysis, our pipeline recomposes the electrogastrogram from all recorded channels after automatic rejection of nuisance components. The main benefits of this approach are reduced attrition, retention of data from all recorded channels, and reduced influence of researcher bias. In addition to our tutorial on the method, we offer a proof-of-principle in which our approach leads to reduced data rejection compared to established methods. We aimed to describe each step in sufficient detail to be implemented in any programming language. In addition, we made an open-source Python package freely available for ease of use.

# Background

Electrogastrography (EGG) is primarily a diagnostic technique, used to identify various disorders of the stomach (Koch & Stern, 2004a). It can also be used to index nausea, e.g. during fullness or motion sickness (Geldof et al., 1986; Koch, 2014; Levine, 2005). In psychological research, electrogastrography can be used to measure proto-nausea (Alladin, Judson, et al., 2024): subtle changes in the electrogastrogram that occur when adults feel disgust towards unappetising food (Stern et al., 2001) or unpleasant stimuli like bodily effluvia (Shenhav & Mendes, 2014; Vianna & Tranel, 2006; Zhou et al., 2004). This could be a learned response, as it seems curiously absent in children (Alladin, Berry, et al., 2024). In addition, electrogastrography is used to characterise brain-body connectivity (Harrison et al., 2010; Rebollo et al., 2018; Richter et al., 2017), and its relationship with mental wellbeing (Banellis et al., 2025).

Despite its benefits, electrogastrography is a difficult technique to implement due to prohibitively expensive hardware and software, and due to idiosyncratic properties of the signal (e.g. the low frequency of the gastric rhythm). Here, we describe a low-cost solution to data acquisition, using the OpenBCI Ganglion amplifier (~$500 at time of writing). We also describe a processing pipeline that retains as much of the gastric signal as possible in noisy conditions.

**What is electrogastrography?**

The enteric nervous system is a branch of the autonomic nervous system, comprising the innervation of the gastric tract and its connections to the central nervous system (Sharkey & Mawe, 2023). In the stomach, the interstitial cells of Cajal form an electrically-coupled network between the corpus and the antrum (Sanders & Ördög, 2003). These cells rhythmically depolarise, activating the voltage-gated $Ca^{2+}$ channels on the smooth muscle cells of the stomach (Sanders & Ördög, 2003; Thomson et al., 1998). This results in a contraction that drives peristalsis. The action potentials in the corpus summate and propagate the ‘electrical halo’ of depolarisation down to the distal antrum (Koch & Stern, 2004b). This results in a contraction that moves down the stomach every ~20 seconds, resulting in the signature 3 cycles per minute (cpm) slow wave rhythm.

The rhythmic properties of the stomach were first discovered and measured by Alvarez (1922) via *in-vivo* electrogastrography, but today they can be measured using non-invasive methods. While invasive electrogastrography is conducted by placing the sensors on the stomach serosa itself (e.g., during surgery), non-invasive electrogastrography is measured by placing cutaneous electrodes over the stomach pacemaker region of the abdominal surface. The continuous recording yields an electrogastrogram, and its frequency decomposition (e.g. using Fournier transform) can be divided into three frequency-related bands: bradygastria (<2 cpm), normogastria (2-4 cpm), and tachygastria (4-9 cpm) (Yin & Chen, 2013). In a regular electrogastrogram, power should be highest in the normogastric band.

### Current aims

The main goal of this tutorial is to describe the acquisition and analysis of electrogastrography, with a focus on low-cost solutions that make electrogastrography more accessible to less well-funded labs. To support this, we demonstrate a standardised processing pipeline that improves data retention (i.e. using all recorded channels rather than choosing the “best” one) and reduces reliance on subjective decisions to reduce potential experimenter bias. We also include an open-source Python package that supports both acquisition and analysis.

## Data Acquisition

### Considerations before recording an electrogastrogram

Electrogastrographic signal is clearest when recorded in a post-prandial state, i.e. within one hour after food consumption. Koch and Stern (2004) outline increased amplitude of the slow waves following food consumption (due to an increased $Ca^{2+}$ channel activation) and a signature 3 cpm rhythm appearance after a balloon inflation in an empty stomach.

The abdominal area should be relaxed to avoid interference of muscle tension with the recording. A supine position is typically recommended (Yin & Chen, 2013), but a reclined position also yields decent data and allows for visual stimuli to be used at the same time (Alladin, Berry, et al., 2024; Wolpert et al., 2020). Any movement (e.g., tensing the abdominal muscles or talking) should be minimised to avoid artifacts.

Prior to sensor application, the skin on the abdomen should be prepared with a gentle exfoliant, and potentially shaved if a participant’s body hair prevents electrode application. These steps ensure low impedance (under 10 kΩ). One to several (typically four) recording electrodes can be used with a common reference electrode, or each recording electrode can have its own reference (bipolar).

The sensors are positioned in the area over and around the stomach pacemaker region. Some studies report using ultrasonography or computer tomography for establishing the position of the stomach individually in each participant (Oczka et al., 2024). However, using anatomical landmarks is less resource-intensive and works well (Rebollo et al., 2018; Wolpert et al., 2020; Yin & Chen, 2013). In four-channel electrogastrography, the reference electrode is placed just below the xiphoid process, and the first recording electrode is placed on the midpoint between the reference and the umbilicus. The remaining electrodes are placed as per Figure 1, with the ground positioned below the participant's left costal margin.

Electrogastrography can also be recorded using a single-channel system, with the recording electrode being positioned over the pacemaker region and the reference to its side (Koch & Stern, 2004b). While multi-channel systems have been reported more often and are argued to provide a better signal-to-noise ratio (Rebollo et al., 2018; Riezzo et al., 2013), single-channel setups can be effective and less resource-intensive (i.e. fewer single-use sensors to be discarded) if proper experimental techniques are ensured (Oczka et al., 2024). A well-positioned single channel ensures good signal and less data processing, but there is a significant chance of data loss due to interpersonal anatomical differences (Oczka et al., 2024). Contrarily, Wolpert et al. (2020) used a 7-electrode system to ensure clear gastric signal from every participant. Ultimately, the choice of the number of channels should depend on the aims of the research and the amplifier that is available.

Finally, electrogastrography recordings should take ideally last no less than 30 minutes, with the clinical gold standard being 60 minutes due to the slow rhythmic activity of the stomach (Levanon et al., 1998; Wolpert et al., 2020).

*Who can take part in an electrogastrography study?*

It is recommended that participants with stomach rhythm abnormalities (e.g., due to Irritable Bowel Syndrome) should be excluded, and that any medications affecting the stomach myoelectrical activity (e.g., proton pump inhibitors or antiemetics) should be stopped at least 1 week before testing (Riezzo et al., 2013). Perhaps obviously, researchers should weigh the importance of their study against the effects on participants of pausing their medication; a comparison that should tilt towards preserving participant wellbeing.

## Electrogastrography recording

To offer a practical example, we outline a procedure to use an OpenBCI Ganglion amplifier for 4-channel electrogastrography. Before sensor placement, the skin surface should be exfoliated (Yin & Chen, 2013). For this example, we used Nuprep (Weaver and Company, Aurora, CO, USA), and used OpenBCI's GUI software to ensure that impedances were below 10 kΩ for all sensors.

Place the reference electrode just below the xiphoid process, and the ground electrode on the participant's left side (just below the costal margin). Then place four recording electrodes: sensor 3 was halfway between the umbilicus and the reference, sensor 4 to the participant's right of sensor 3, sensor 2 up and to the participant's left, and sensor 1 in the same direction (i.e. 45°) from sensor 2. The distance between sensors is typically 3-5 cm. Sensor placement is depicted in Figure 1.

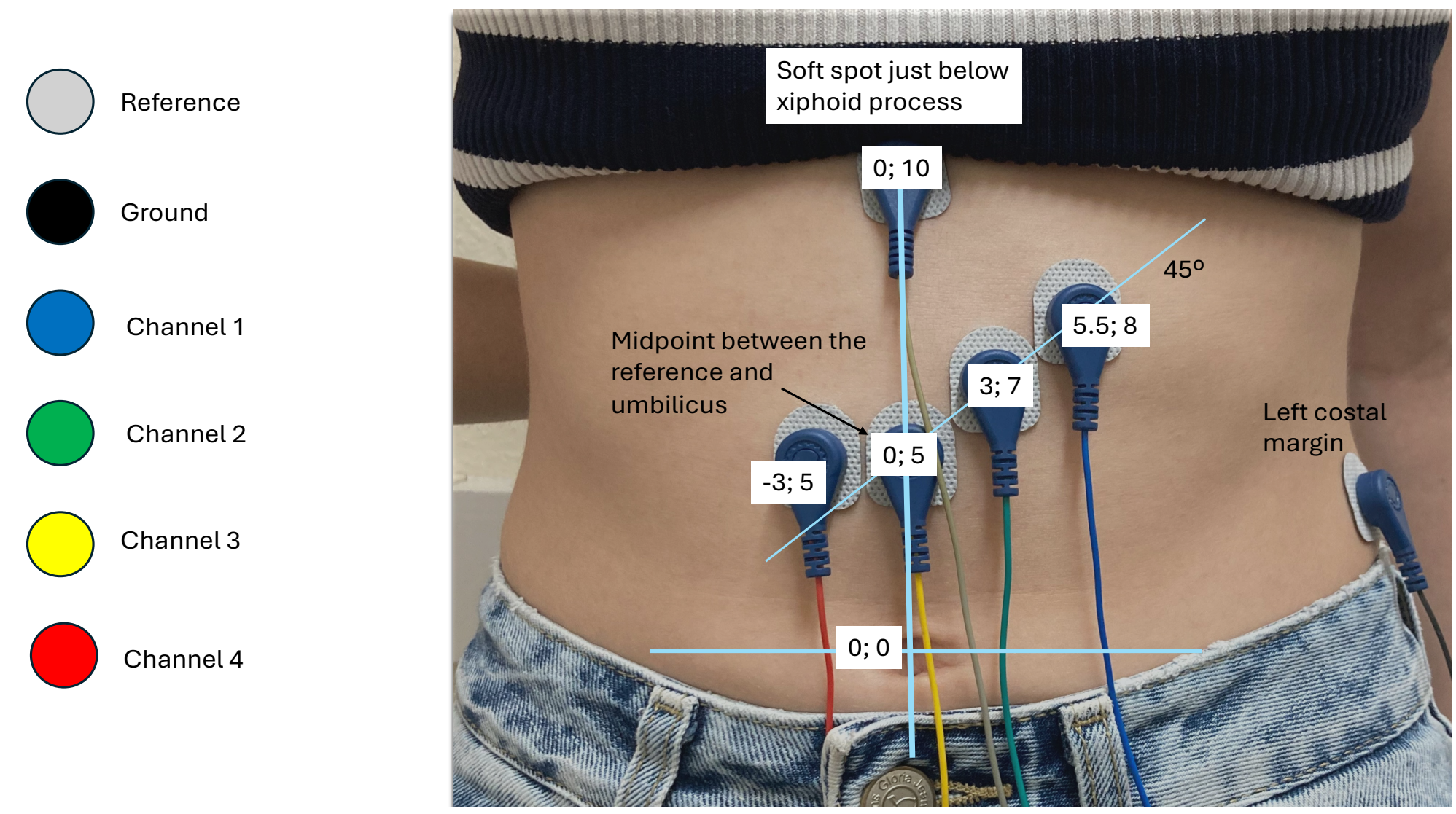

***Figure 1****. Sensor placement for eletrogastrography, with annotations to indicate anatomical landmarks and sensor coordinates. Position are measured in centimetres, using the umbilicus as the origin. The reference (with grey cable) is placed just below the xiphoid process, and the ground (black) on the (participant) left side just below the costal margin. Recording sensors are placed in the above pattern: sensor 3 (yellow) is halfway between reference and umbilicus, sensor 4 (red) is to the participant's right of sensor 3, sensor 2 (green) is equal parts up and to the participant's left from sensor 3, and sensor 1 is on the same 45° line from sensor 3 as sensor 2. Distance between recording pictured recording sensors is ~3 cm, and ~3-5 cm for all participants.*

In our lab, we use 20x25 mm disposable Ag-AgCl cutaneous electrodes with a snap connection (Spes Medica, Neurotab, DENIS0205). Our participants are sat in a reclined position during recordings, using Ikea “POÄNG” armchairs (available in child and adult sizes).

While electrogastrography is not an intimate procedure, we do ensure that a chaperone is present during sensor placement and removal. We also allow participants to indicate their preference for experimenter gender, and offer the option to bring a trusted individual.

## What happens after data acquisition?

Following data acquisition, several filtering steps are employed to isolate the gastric signal. Pre-processing steps vary across literature. For example, the electrogastrogram can be visually inspected for motion artifacts and then processed through a spectral analysis (Levanon et al., 1998; Yin & Chen, 2013), and sometimes a fast Fourier transform is applied to the data prior to the visual analysis (Riezzo et al., 2013). Recent literature suggests a more comprehensive pre-processing pipeline, including a power spectrum calculation followed by a visual analysis, with subsequent filtering and calculating whether most cycles lie within the normogastric range (Wolpert et al., 2020). Most published electrogastrography pre-processing pipelines rely on such visual inspections of the data, leading to potential experimenter bias and data loss. Fully automated procedures have recently been introduced (Alladin, Berry, et al., 2024; Gumussoy et al., 2025), and are described in the *Data Analysis* section of this tutorial.

## Suitable low-cost hardware for electrogastrography acquisition

OpenBCI is a USA-based producer and retailer of amplifiers (“boards”) intended for use in human electrophysiology by professionals and consumers. The company is popular among low-resource labs because its hardware is comparatively low-cost and its software for data acquisition and analysis is open-source (Ha et al., 2023). The 8-channel OpenBCI Cyton amplifier has been used with custom hardware for cutaneous (non-invasive) and serosal (invasive) electrogastrography in pig (*Sus domesticus*) (Sukasem et al., 2022), and for mobile electrogastrography in humans (Gharibans et al., 2018; Vujic et al., 2020).

Here, we describe data from a 4-channel OpenBCI Ganglion board, which is substantially less costly at about half the price of Cyton at the time of writing. The Ganglion is a battery-operated amplifier that can record from up to four channels at 200 Hz, and streams data via Bluetooth or to an on-board SD card. It has previously been used for electroencephalography (Auda et al., 2022) and electromyography (Cao et al., 2019; Peterson et al., 2020).

## Code example for data acquisition

Equipment differs between labs, and it would be challenging to provide a comprehensive overview of computer code for all possible devices. To ensure brevity but also inclusion of less-resourced labs, we decided to focus on a relatively low-cost option. This comprises the hardware described above (OpenBCI Ganglion) and a freely available Python package for recording data using boards from this manufacturer. This package offers a simplified interface for the BrainFlow API (www.brainflow.org).

### *Installing the Python package for acquisition and analysis*

Our Python packages has been made available through the Python Package Index (PyPI), and can be found on https://pypi.org/project/electrography/. The source code has also been made available on GitHub: https://github.com/esdalmaijer/electrography. The easiest way to install the package is by using Python's package installer pip. You can do so by opening a terminal (Linux and macOS) or a Command Prompt (Windows), and running the following command:

```
python -m pip install electrography
```

***Listing 1***. *Installing our Python package can be done in a Terminal or Command Prompt, using this code. The alias "python" can also be "python3" or "py" on different operating systems.*

If you work with Python environments, make sure to activate the relevant environment before running the installation command. If you use the Anaconda distribution, ensure that you are running its pip (e.g. on Windows this requires using the Anaconda Command Prompt rather than the default Command Prompt).

### *Importing the OpenBCI class*

To use the OpenBCI board interfaces included in our package, they have to be imported into an experiment. To do so, use the following code among your other imports (usually at the top of a script):

```
from electrography.boards import OpenBCI
```

***Listing 2***. *Code for importing the OpenBCI board interface.*

*Opening a connection*

Once the OpenBCI class is imported, an instance needs to be created to run a recording session. This requires the board's name, data specific to the type of board, and a file path for data to be written to. For Ganglion boards, this requires a serial port address, e.g. "COM4" on Windows or "/dev/ttyACM0" on Linux. The file path used in the example below is simply a single file name, which will produce a file in the active directory. A more sensible approach is to provide a full path, e.g. to an intended file in a data directory.

```
board = OpenBCI("ganglion", serial_port="/dev/ttyACM0", \
    log_file="example_egg.tsv")
```

***Listing 3***. *Code to initialise an OpenBCI Ganglion board connection. Note that the serial port address can be different on your operating system and number of connected devices.*

*Starting and stopping the recording of data*

Data can be streamed from the amplifier to a file. This process is not automatically started upon the successful initialisation of an OpenBCI instance, but has to be started using the function below. We recommend doing so early, to allow the amplifier to warm up and connection to settle in. (This is a recommendation based on anecdote, as we occasionally see low-frequency high-amplitude artifacts at connection onset in our lab.)

```
board.start_stream()
```

***Listing 4***. *Code to start the recording of data to a file.*

The recording of data to file can be paused by using the function listed below. In many other scenario's, one could do so to pause the recording outside of trials and blocks. However, electrogastrography requires long uninterrupted recordings, so we recommend against frequent pauses. We typically record throughout a session, and extract relevant data afterwards.

```
board.stop_stream()
```

***Listing 5**. Code to pause the streaming of data to a file. This will not close the connection, but it will the streaming of data to disk.*

*Recording events in the data*

In order to make sense of data after recording, one needs to know which data correspond to which event in the experiment. For example, when recording electrogastrography during blocks corresponding to experimental conditions, it needs to be known when those blocks started and finished, and which condition they reflected. To ensure this is possible, it is common practice to record markers in the data when specific events (e.g. the start of a block) occur.

In higher-frequency data (e.g. electroencephalography), it is common to provide external input to an amplifier to record experimental events. This can take the form of a photodiode that detects changes on the screen and marks them in a data file. In addition, markers are often written to the data file by experimental software to code when events happen, but also what those events are (e.g. to differentiate between stimuli or types of trials).

Traditionally, as a consequence of mostly historical limitations on bandwidth, event markers are coded as an 8-bit unsigned integer. This means they can be 0 (no event), or any value between 1 and 255 (inclusive). How events are coded is up to the researcher. For example, in an experiment gauging the perception of individual pop singers, they could code stimulus onset 11 for one singer, 12 for the next, and 13 for Taylor Swift.

```
board.trigger(250)
```

***Listing 5.** Code to include an event marker in the data file, which can be used to identify relevant data for analysis. In this case, the value 250 is used.*

*Closing a connection*

Once an experiment has finished or enough data has been collected, the connection between computer and amplifier needs to be closed. This ensures that data files are closed properly, and that the amplifier is left in a state in which it is ready for the next recording. To do so, one needs to call the OpenBCI instance's close function.

```
board.close()
```

***Listing 6.*** *Code to end communications with the amplifier. This neatly closes the file and the connection to the board.*

## Data Analysis

### Analytical pipeline

The proposed analytical pipeline is described below, per step. Many of these are existing pre-processing steps in electrogastrography research, combined with a number of novel approaches that aim to retain data, reduce noise, and limit subjective researcher decisions. Code listings are provided for each of the steps. These showcase functions from the open-source Python package that we have made available alongside this manuscript.

Every step is illustrated with 5 minutes of data from an example recording, visualised in original and in frequency domain. The file for this recording, "example_egg.tsv", has been included as supplementary material to this manuscript. Readers are encouraged to download this to play along at home while reading through the following code listings.

*Step 0: loading Python modules*

Before being able to use the functions in the listings below, they need to be imported from the electrography module. The NumPy package should also be loaded.

```
import numpy
from electrography.data import read_file, get_physiology, mad_filter, \
    butter_bandpass_filter, movement_filter, ica_denoise
```

***Listing 7.*** *Code example in which the relevant functions are loaded (in order of appearance in this tutorial).*

*Step 1: load data*

We provide an example that assumes a recording was conducted with an OpenBCI amplifier using our Python package. While the code example is specific to this format, the universal element is that data is loaded into an (M,N)-shaped array, with M being the number of channels and N the number of recorded samples.

```
file_name = "example_egg.tsv"
raw = read_file(file_name)
data, t, triggers, sampling_rate = get_physiology(raw, "ganglion")
```

***Listing 8***. *Code example to load data from a recording stored in "example.csv", and to extract the relevant physiological data, timestamps (in ms), triggers (codes logged when events occurred in the experiment), and sampling rate. The variable* data *is a NumPy array of shape (M,N) where M is the number of channels and N is the number of recorded samples.*

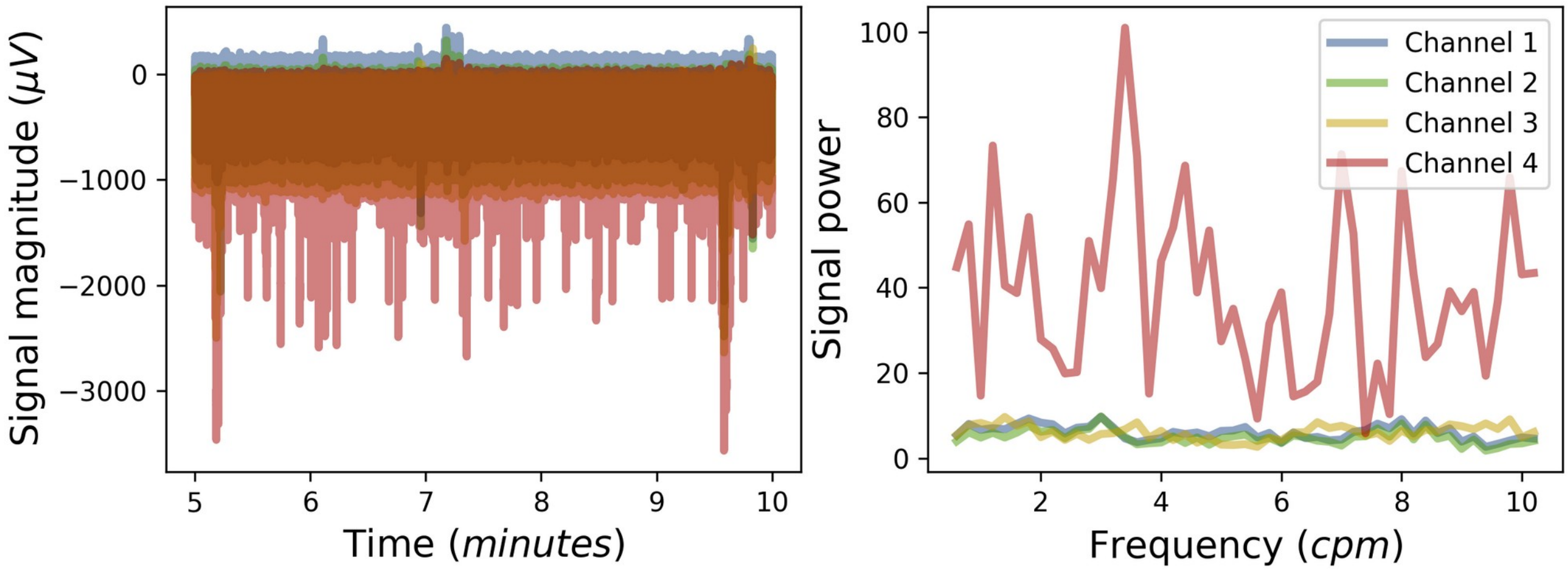

***Figure 2.*** *Five minutes of data from four electrogastrography leads (see Figure 1) plotted as a timeseries (left) and in the frequency domain (right) after loading the data in step 1. The signal shows as an undifferentiated band in original space, as it captures high frequency deflections associated with the heart's sinus rhythm. Also apparent is that deflections on channel 4 (red) are of a much higher amplitude than the other leads, which suggest they might reflect artefacts.*

*Step 2: mean-centring*

Not all signal is centred with a mean of zero. When subjected to e.g. fast Fournier transform, such offsets will appear as a constant at 0 Hz in the frequency domain. This could look like or obscure genuine movement artifacts, which typically occur at low frequencies. To reduce this, one could subtract the mean of a signal from each of its samples.

```
m_per_channel = numpy.nanmean(data, axis=1, keepdims=True)
data = data - m_per_channel
```

***Listing 9***. *In this code example, the mean is computer for all channels (using the* nanmean *function to ignore any missing data), and then subtracted from that signal.*

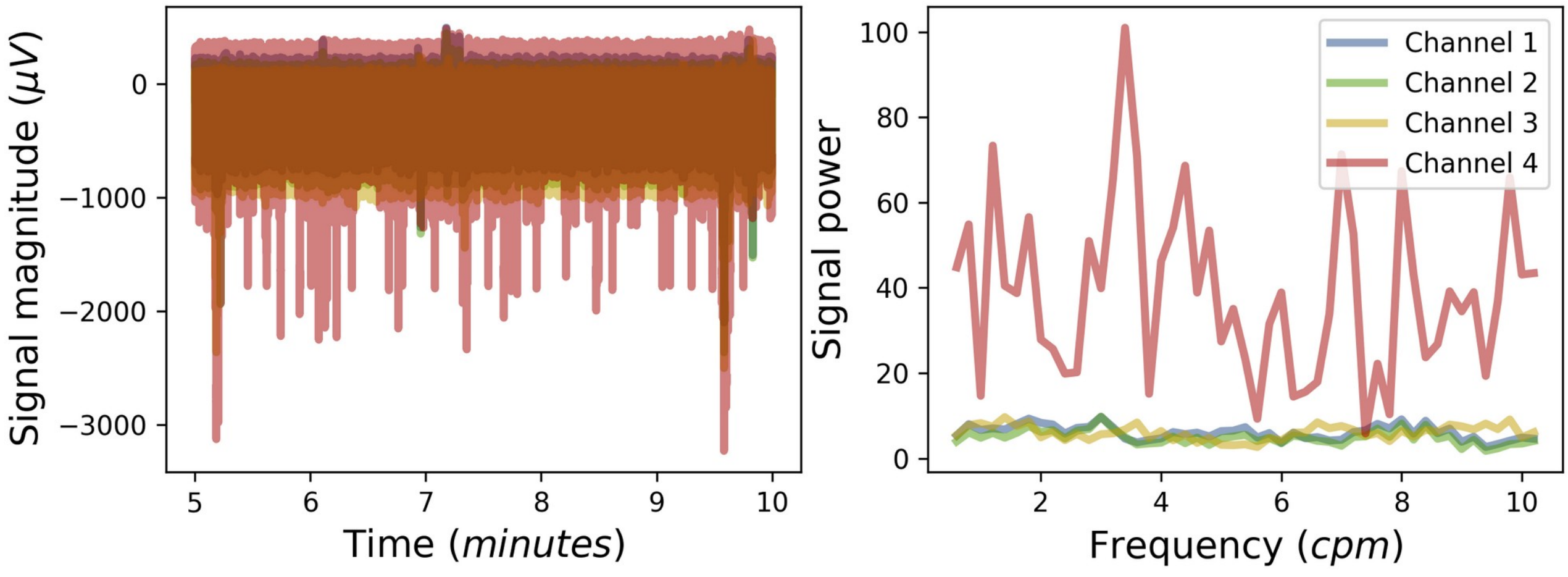

***Figure 3.*** *Five minutes of data from four electrogastrography leads (see Figure 1) plotted as a timeseries (left) and in the frequency domain (right) after mean-centring the data in step 2. This changes very little about the data profile, but each lead's average is now at 0 μV.*

*Step 3: outlier rejection*

Electrophysiological recordings are sensitive to several sources of noise that show up as spikes (impulse transients), such as cable or sensor tugging, static discharge from clothing, or a gardener using heavy equipment directly outside of an under-insulated lab. These transients often appear as high-amplitude voltage spikes that can be easily spotted in a plot of the data. Due to their magnitude, spikes are likely to impact any downstream analysis, and should thus be filtered out.

Outlier rejection can be done through a Hampel filter. This computes the median and median absolute deviation in a running window, and replaces all values that fall outside an acceptable boundary by the median. As gastric signal is a slow wave, the Hampel window length could be wider than typical. In practical terms: the default window length in Matlab's *hampel* function is 7 samples; but a single normogastric wavelength is 20 seconds, which at 200 Hz translates to 4000 samples. The benefit of a wider window is better filtering of spikes that span across several samples. The cost is increased processing time.

```
data = hampel_filter(data, k=2000, n_sigma=3)
```

***Listing 10****. In this code, a Hampel filter is applied to remove outliers from the data in each channel. The data is passed as a multi-dimensional array with all channels, and the filter is applied to each channel in parallel (i.e. simultaneously but independently). The window size is set by parameter* k*, which sets its extent in both directions from the current sample, here resulting in a total window length of 4001. The parameter* n_sigma *sets the outlier threshold, here 3 times the standard deviation as estimated from the median absolute deviation.*

Due to their iterative nature, Hampel filters are computationally expensive. This is especially true when windows are wide. If signal shows no or low drift, there is a substantially reduced need for a running window. In this case, one could opt for computing a single median and median absolute deviation (MAD) over the whole signal, which is much faster. Note that the function, *mad_filter*, is named after the abbreviation and does not necessarily reflect the common consensus on its usage.

```
data = mad_filter(data, n_sigma=3)
```

***Listing 11****. In this code, the mad_filter function computes a median for each channel in the signal, using the* nanmedian *function to ignore any missing values. The absolute deviation and its median are then computed, again for each channel separately. Then, a standard deviation is estimated from the median absolute deviation by scaling it with* $\kappa \approx 1.4826$ *(the reciprocal of the product of the square root of 2 and the reciprocal error function of a half), as is typical in Hampel filters. Finally, samples for which the absolute deviation exceeds the threshold (here set as 3 standard deviation equivalents) are replaced by the channel median.*

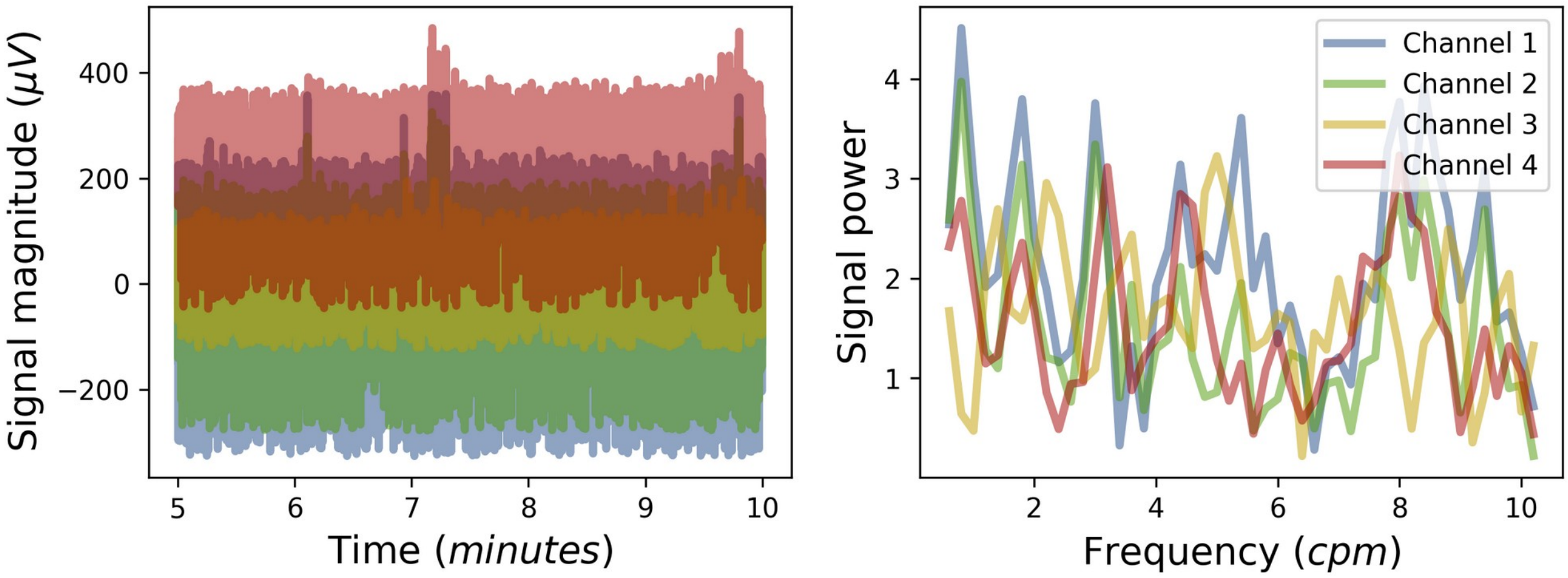

***Figure 4.*** *Five minutes of data from four electrogastrography leads (see Figure 1) plotted as a timeseries (left) and in the frequency domain (right) after outlier rejection with a mean-absolute-deviation filter in step 3. The outlier rejection is most notable in channel 4 (red), where the high-amplitude artefacts are no longer apparent.*

*Step 4: frequency filtering*

In electrogastrography, the signal of interest is between the frequencies of 0.5 and 10 cycles/minute (roughly 0.0083 to 0.17 Hz). Signal outside of these frequencies should thus be filtered out. There are several frequency filters that could accomplish this, and here we demonstrate a Butterworth filter. It has a high-pass (higher frequencies are allowed through) of 0.5 and a low-pass (lower frequencies are allowed through) of 10 cycles/minute, so the band between 0.5 and 10 cycles/minute is filtered in.

Frequency filters can shift data in time. This is not necessarily an issue in the frequency domain across a recording period. However, if the data is used in time-sensitive analyses (e.g. phase-amplitude coupling between electrogastrogram and blood oxygenation measured through magnetic resonance imaging), time shifts should be addressed. A common way to do this is to run the same frequency filter on the reversed data (addressed here by using the bandpass filter function’s “bidirectional” keyword).

```
for channel in range(data.shape[0]):
    data[channel,:] = butter_bandpass_filter(data[channel,:], \
        0.0083, 0.17, sampling_rate, bidirectional=True)
```

***Listing 12**. In this code, a Butterworth filter is applied to each channel. Its lower bound is 0.5 cycles per minute (0.0083 Hz), its high bound is 10 cycles per minute (0.17 Hz), and its sampling rate was defined in Listing 1 (200 Hz in our example).*

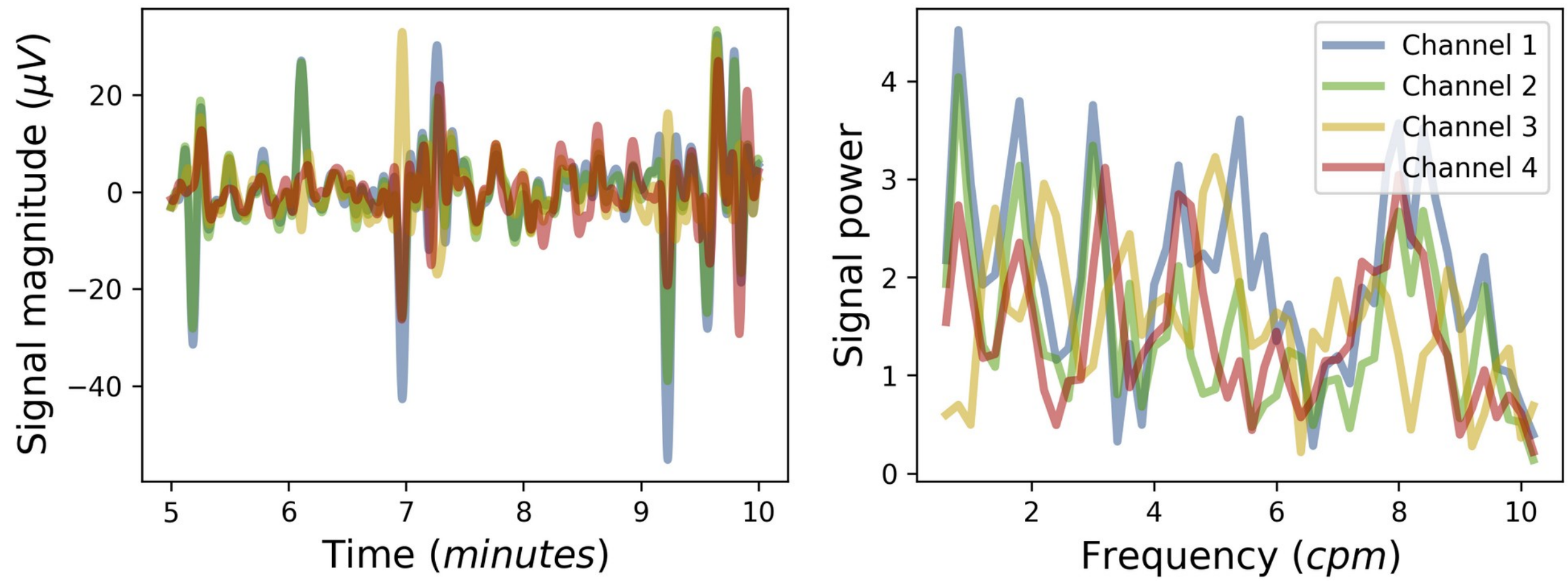

***Figure 5.** Five minutes of data from four electrogastrography leads (see Figure 1) plotted as a timeseries (left) and in the frequency domain (right) after a bandpass filter was applied in step 4. The frequency filter has excluded all non-gastric frequencies, which has made a gastric-like rhythm more apparent in the timeseries. However, the characteristic peak at 3 cycles per minute is still not particularly evident in the frequency domain.*

*Step 5: movement rejection*

Participants move during testing. While instructions not to fidget can work (to some extent), it would be unreasonable to expect participants to not move (e.g. to readjust) during the long recording periods that electrogastrography requires. Such movements impact recordings in two ways: as high-frequency (up to 100 Hz) noise from muscle activity, and as low-frequency noise through the movement itself. The electromyographic noise will have been filtered out in the previous step, as it occurs well above the gastric band (Mills, 2005). Lower-frequency noise can be tackled using a movement filter designed for ambulatory electrogastrography (Gharibans et al., 2018). Another benefit of this technique is that it can filter out oscillatory transients, which can also appear as high-amplitude low-frequency artifacts.

The Gharibans movement filter aims to predict noise from the electrogastrogram, so that it can be subtracted from the recorded signal. It uses the fact that the main normogastric frequency is known (3 cycles/minute; 0.05 Hz), and that movement typically appears at a much higher amplitude than gastric signal. It comprises a linear minimum mean-squared error estimator with a window size of one normogastric wavelength, and uses local variance to estimate the artifact signal.

Because of the iterative nature of the sliding window, the movement function is computationally expensive. We thus implemented the filter in a statically typed function (compiled using Cython) for better speed, and a fallback that is slightly less quick (using NumPy).

```
data, noise = movement_filter(data, sampling_rate, 0.05, window=1.0)
```

***Listing 13****. This code applies the Gharibans et al. movement filter. It uses a statically typed function if a suitable compilation is available, and otherwise falls back on a NumPy implementation that is slower but results in the same outcome. The sampling rate is defined in Listing 1. The next parameter sets the frequency of interest, which is 0.05 Hz (3 cycles/minute) for gastric signal. The* window *parameter sets the window length in cycles of the frequency of interest, and should ideally be kept at 1 for most operations. The function returns both the data after filtering out movement noise (here captured by the data variable), and the estimated movement noise (here captured by the* noise *variable).*

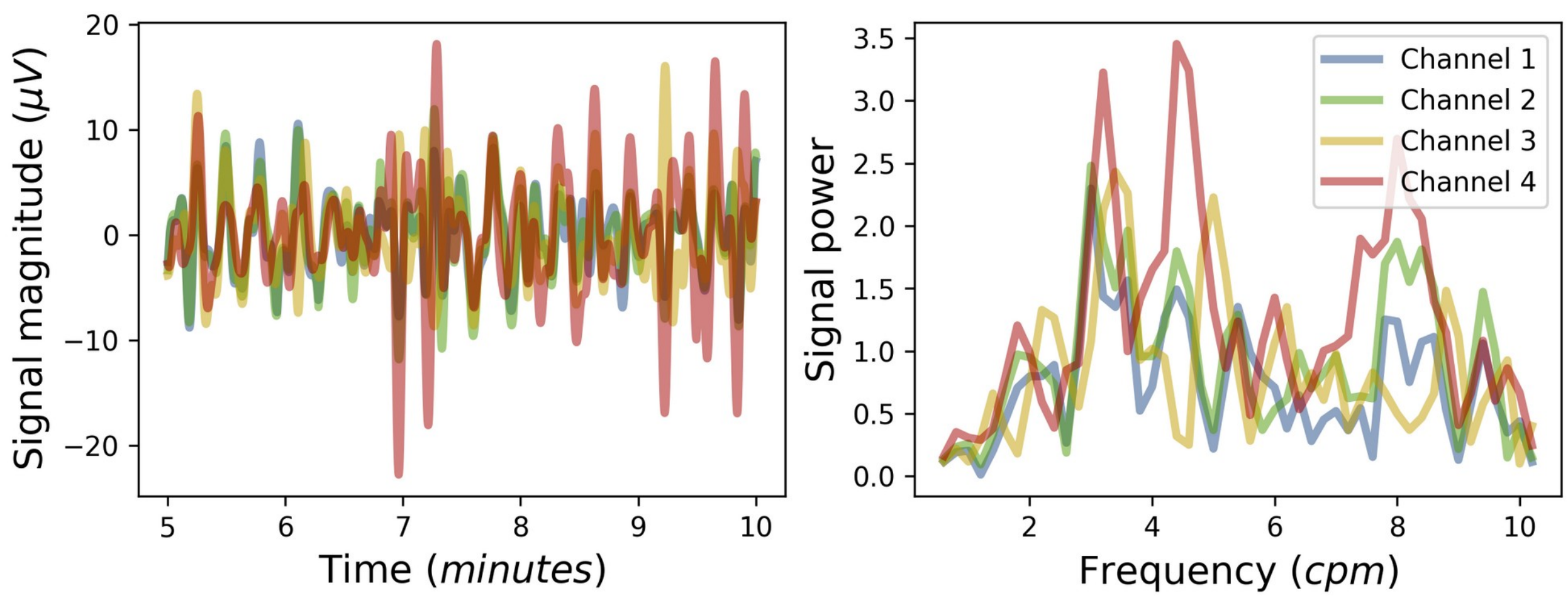

***Figure 6.*** *Five minutes of data from four electrogastrography leads (see Figure 1) plotted as a timeseries (left) and in the frequency domain (right) after movement was filtered out of the data using the approach from Gharibans and colleagues (2018) in step 5. After this step, the gastric peak has become much clearer at its expected frequency of 3 cycles per minute. Other peaks remain apparent, including at 4-6 and 8-9 cycles per minute.*

*Step 6: signal isolation and de-noising*

In clinical applications, multi-channel recordings could be used to diagnose abnormalities in gastric wave propagation (Gharibans et al., 2017). Instead in research, a common approach to multi-channel electrogastrography is to manually identify the “best” channel, and only use this for further analysis (Banellis et al., 2025; Rebollo et al., 2018; Wolpert et al., 2020). While this approach is the de facto gold standard and works well to identify gastric signals in a majority of participants, it leaves some room for improvement. For example, the current approach is explicitly subjective, excludes participants with noisy or otherwise imperfect recordings, and discards information from unselected channels.

Here, we suggest an approach that aims to identify and isolate the sources underlying the electrogastrogram, removing any sources of noise and non-gastric activity. To do this, we use independent component analysis on all recorded channels. This decomposes the recorded signal into the sources that likely made it up. We then use fast Fournier transform on each of the components, and compute the peak power in the normogastric range of 2-4 cycles/minute (the “signal”) and the average power across the remaining spectrum of 0.5-2 and 2-10 cycles/minute (the “noise”). This allows us to compute a gastric signal-to-noise ratio for each component. Only components with a ratio of 3 and over are retained, and all others are zeroed out. This threshold is consistent with the 3:1 standard for using electrophysiologic equipment in neurological settings (Cadwell & Villarreal, 2012). Finally, the source components are inverse-transformed into their original space, leaving a de-noised multi-channel gastric recording.

```
data = ica_denoise(data, sampling_rate, 0.0083, 0.17, [0.033, 0.067], \
    snr_threshold=3.0, random_state=1989)
```

***Listing 14**. This code uses an independent component analysis (ICA) on the data to estimate its potential sources. It then computes the signal magnitude across frequencies (here between 0.5-10 cpm, or roughly 0.0083-0.17 Hz) for each component independently, and then the signal-to-noise ratio as the ratio between the normogastric peak power (2-4 cpm, which is roughly 0.033-0.067 Hz) and the average power outside of it (i.e. in the brady- and tachygastric ranges). Any componentwith a signal-to-noise ratio under the threshold (here set to 3) is zeroed out. Afterwards, the components are inverse-transformed back into original signal space.*

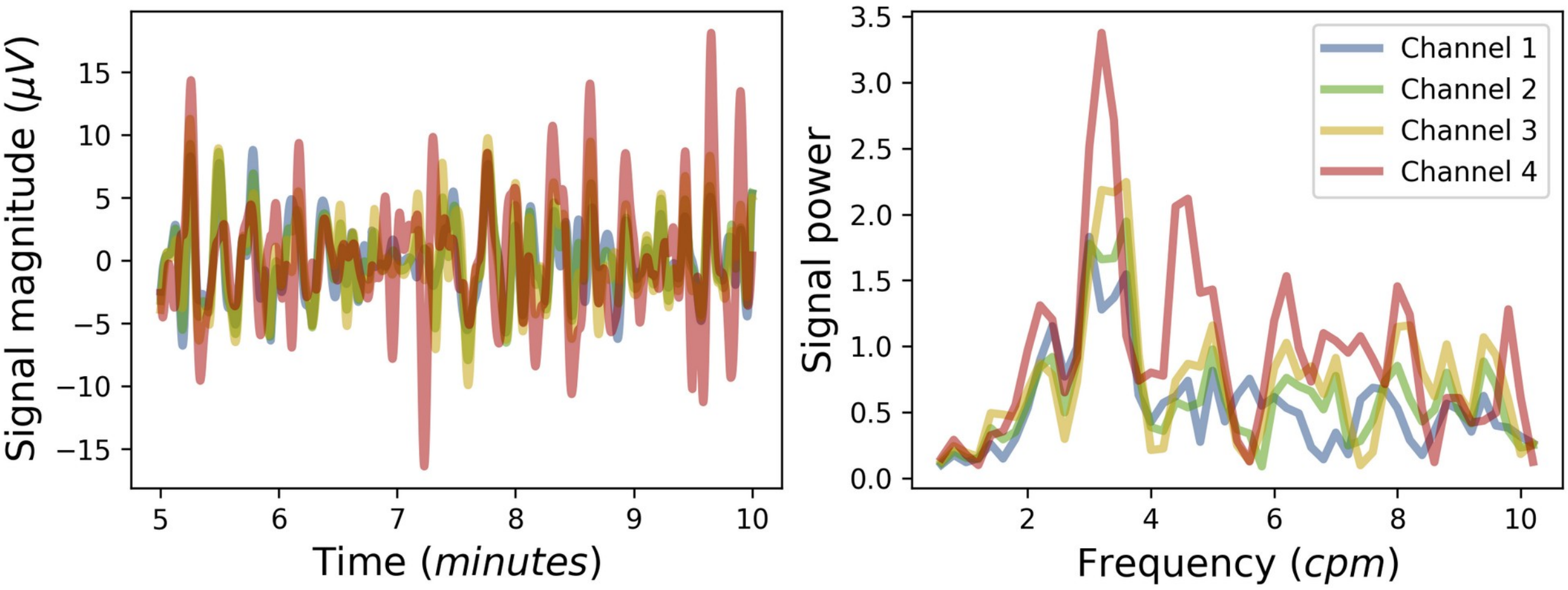

***Figure 7.*** *Five minutes of data from four electrogastrography leads (see Figure 1) plotted as a timeseries (left) and in the frequency domain (right) after denoising the data using an ICA decomposition in step 6. The secondary peaks at 4-6 and 8-9 cycles per minute that remained after step 5 are no longer apparent. This were identified as a non-gastric component in the independent component analysis, and thus removed before the signal was reconstituted.*

For an analogy of this technique, think of microphones spread around a stage during a music performance. Each microphone records a unique combination of all performers: singer, guitarist, bassist, and drummer. Independent component analysis uses subtle differences between the microphones to separate them into constituent sources. If we only care about the vocals, we could zero out all components that do not sound like singing (i.e. those capturing guitars and drums). After doing this, the reconstructed signal would be the isolated vocals, with individual channels reflecting how each microphone picked up the singer (i.e. louder for those closer to them).

In the case of electrogastrography, there are no microphones conveniently located near the pacemaker region. Hence, the analogous situation would be that microphones having been scattered across the stage, which necessitates averaging signal across microphones as all could have recorded the vocals equally well.

One may be tempted to forego inverse transformation, and simply choose the independent component with the best gastric signal-to-noise ratio. However, gastric signal is not always isolated in a single component. Hence, extracting only one risks accidentally excluding meaningful data. For example, a participant's typical gastric rhythm could be captured in one component, while their response to an experimental condition that induces proto-nausea could be captured in another. In this case, including only the stereotypical gastric component would have erased the conditional effect.

*Step 7: further analysis*

After the final step of signal isolation and de-noising, the data is fully pre-processed. Further analysis can be as simple as casting the data in the frequency domain to assess the relative power in brady-, normo-, and tachygastric bands (Yin & Chen, 2013), or the differences in gastric profiles between experimental conditions (Alladin, Berry, et al., 2024). It could also be used to compute phase-amplitude coupling between electrogastrogram and magnetoencephalogram (Richter et al., 2017), or phase-locking with the blood-oxygenation-level-dependent response from functional magnetic resonance imaging (Rebollo et al., 2018).

# Pipeline Validation

The aim of this validation was to test if our proposed method had benefits over the “select the best channel” approach. We do so by independently testing the Hampel filter and ICA de-noising, and comparing these approaches to the currently established practice of visual inspection.

One downside of selecting a single “best” channel is that data loss is high: for example, about 12% of data or 12-32% of participants in recent electrogastrography papers (Banellis et al., 2025; Rebollo et al., 2018; Richter et al., 2017). We think that our method will reduce this dropout because it is designed to isolate the gastric signal in multi-channel recordings. Our outcome measure will thus be the number of participants retained in the analysis.

*Visual inspection and best-channel selection*

The established approach involves visual inspection of gastric data, and manually selecting the “best” channel. We followed a typical approach of using two raters who each identified the channel they believed best represented the gastric signal (Banellis et al., 2025). We did so in two pipelines: 1) zero-centring, Butterworth filtering, movement filtering, and visual best-channel selection; and 2) zero-centring, Hampel filtering, Butterworth filtering, movement filtering, and visual best-channel selection.

*Filtering and ICA de-noising*

Our proposed approach involves outlier rejection and signal isolation using independent component analysis. Here, we test these steps in two pipelines: 1) zero-centring, Butterworth filtering, movement filtering, and ICA de-noising; and 2) zero-centring, Hampel filtering, Butterworth filtering, movement filtering, and ICA de-noising.

## Validation experiment

The data used to validate our pipeline was collected as part of an undergraduate dissertation project. This is an ideal dataset for our purposes: it was collected by relatively inexperienced students, and thus potentially noisier than data collected by highly trained professionals.

*Participants*

We recruited 32 participants between March and August 2023, most of whom were local university students (19 women, 13 men; mean age 22 years with a standard deviation of 2.9; 26 identified as White, 2 as Middle Eastern or North African, 2 as Black, and 2 as Asian). They were compensated at a rate of £10 per hour. Participants were advised to have a meal no less than 2 hours prior to testing, and only one participant did not adhere. This study was approved by the University of Bristol's School of Psychological Science Research Ethics Committee (reference 15618).

*Procedure*

Participants were informed about the study's procedure and goals, offered the opportunity to ask any questions, and provided written informed consent. They were prepared for electrogastrography (see under "*Electrogastrography acquisition*"). We streamed live data so that participants could see an example trace. Participants were encouraged to tense their abdominal muscles and then to move around, so that they can observe the change in signal. We included this demonstration of movement artifacts to stress the importance of avoiding movement during data acquisition.

After setup, participants watched a passive-viewing task that lasted about 25 minutes. The task entailed two separate presentations of 28 static images that were on screen for 20 seconds each, separated by 7-9 seconds of inter-trial interval. Participants were seated in a reclined position to improve recording quality while ensuring they could watch a desk-positioned monitor (Alladin, Berry, et al., 2024). The experiment was coded in Python 3.8.10 (for a tutorial, see: Dalmaijer et al., 2025), using PyGaze 0.7.3 (Dalmaijer et al., 2014) and its PyGame (v. 2.0.3) back-end.

*Visual inspection of electrogastrography*

The four channels were inspected independently by two researchers to (1) determine if to include or exclude that participant's data, and then (2) select a single channel deemed to reflect the normogastric signal. Before rating, the two individuals discussed what they were looking for – a clear gastric peak in the range 2.0 to 4.0 cpm, with the gastric signal being distinct from surrounding noise. This includes considering signal amplitude, alongside the general trend of signal to noise. Additionally, the two raters discussed considering a clear signal in relation to both the presence of a qualitatively clear peak and the quantitative amplitude values (indicated on the y-axis). Raters then spent approximately 30 minutes independently reviewing each participant's data, followed by a discussion of their ratings.

**Visual inspection methods suffered high attrition**

For the traditional pipeline (zero-centring, Butterworth filtering, movement filtering, and visual best-channel selection), inter-rater agreement was at 85% and data from 19 participants (58%) were included. Data from 14 participants (42%) were rejected. When a Hampel filter between zero-centring and Butterworth filtering was used, one additional participant was included but inter-rater agreement fell to 76%.

The above suggests that substantial data loss occurred during visual inspection. Our rejection rates were somewhat higher than published work. This aligned with our design, as we explicitly aimed to achieve poorer data quality by using less experienced testers to allow for a better test case of our automated methods.

**Automated EGG pipeline reduced attrition at the cost of noisier averages**

The automated pipeline comprised zero-centring, MAD filtering, Butterworth filtering, movement filtering, and ICA de-noising. With the de-noising signal-to-noise threshold set to 3, data from all 33 participants were included.

Visual inspection suggested that clear gastric profiles could identified for many of the previously excluded participants (see e.g. participants 7, 13, 27, and 33 in the Supplementary Materials). However, for other participants the included data seemed noisier than ideal (see e.g. participants 5, 22, and 30 in the Supplementary Materials). In addition, the automated signal isolation process made signal quality worse for some participants (see e.g. participants 18 and 28 in the Supplementary Materials); although it improved it for others (see e.g. participants 20, 23, and 29 in the Supplementary Materials).

The inclusion of more participants resulted in a noisier group average (Figure 8), i.e. with a lower ratio between normogasric peak power and the average power outside of the normogastric band. Group averages became less noisy when the signal-to-noise threshold (the ratio below which independent components were rejected) was increased.

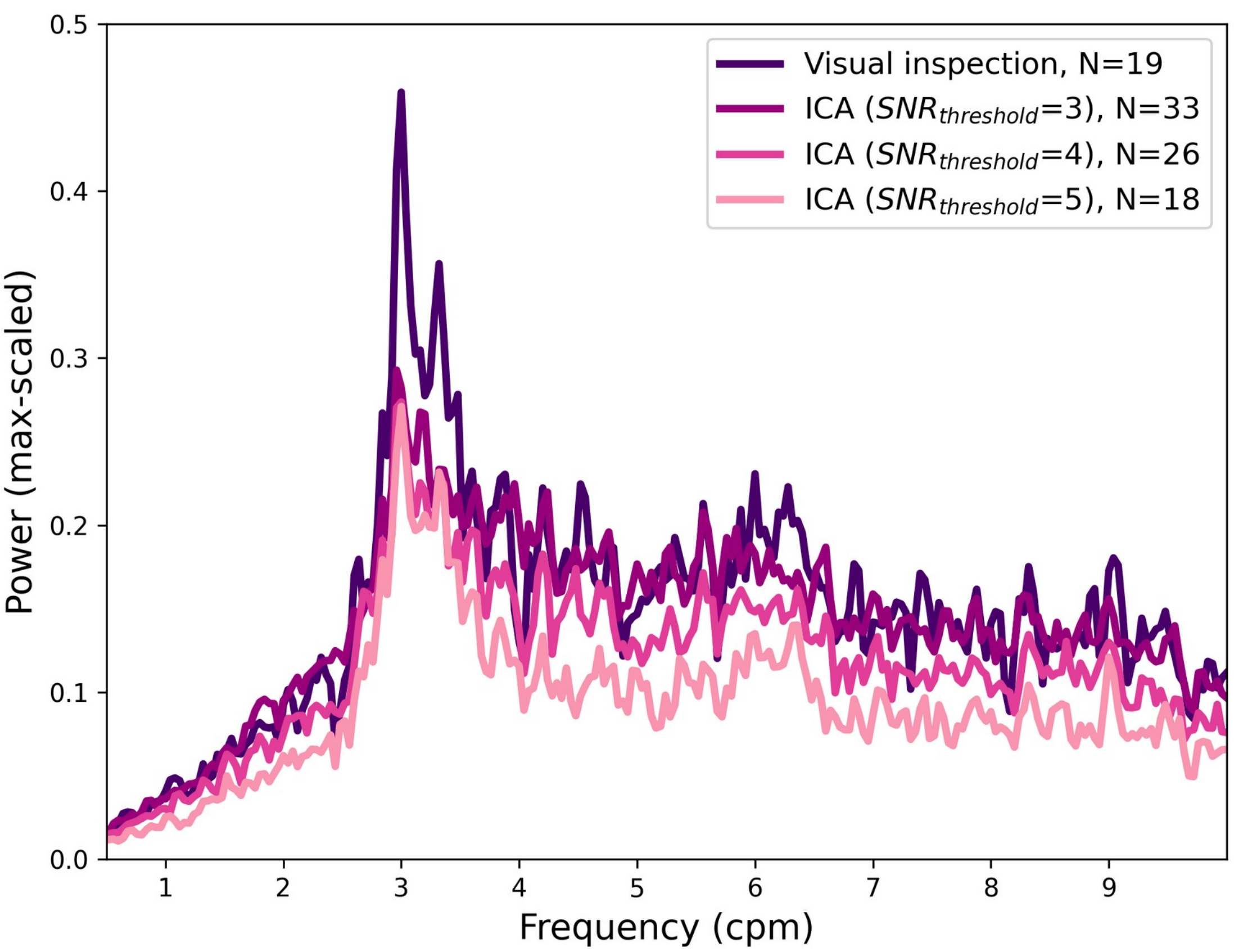

***Figure 8.*** *Gastric power in the frequency domain. Gastric power was computed using fast Fournier transform, and scaled to each participant's maximum (this ensures visual comparability between individuals). The dark purple line reflects data included after visual inspection. The other lines represent data following the pipeline described in this tutorial. The included N for each approach is listed in the legend (maximum N=33). As the signal-to-noise threshold within our pipeline's ICA denoising step increased, noise was reduced but attrition was increased.*

# Conclusion

The electrogastrogram is a slow signal that comes with unique challenges in its acquisition and analysis. In this tutorial, we described what general issues to beware of, including the positioning of participants, recording duration, and sensor placement. We also described a low-cost method for data acquisition using an off-the-shelf amplifier and readily available electrodes, paired with open-source software. Finally, we outlined and compared two methods for EGG analysis. The first is the traditional method of visually inspecting all channels and choosing the “best” gastric signal. We also described an automated signal processing method that includes mean-centring, median filtering (e.g. through a Hampel or a non-windowed median absolute deviation filter), bandpass filtering (here through Butterworth), movement filtering, and finally ICA denoising. The benefit of visual inspection is that the resulting data is of high quality, but this comes at the cost of high attrition. The benefits of the described automated EGG pipeline is that it is less labour intensive and reduces the potential for human bias, that it allows for the combination of data from all recorded channels, and that it substantially reduces attrition. However, this comes at the cost of including more noisy (lower signal-to-noise ratio) datasets. To choose an approach, researchers should weigh the relative benefits of each method against their project’s priorities.